\documentclass[twocolumn]{aastex63}
\graphicspath{{./fig/}{./png/}}
\usepackage{tabularx}
\usepackage{graphicx,graphics,amsmath}
\usepackage{natbib}
\usepackage{bm,url}
\usepackage{times}
\usepackage{amssymb}
\usepackage{color}
\usepackage{float}
\usepackage{arydshln}
\usepackage{multirow}
\usepackage{xcolor}
\usepackage{ulem}
\usepackage{color}

\def\bl{Babcock--Leighton}
\newcommand{\Fig}[1]{Figure~\ref{#1}}
\newcommand{\Eq}[1]{Equation~(\ref{#1})}
\newcommand{\Sec}[1]{Section~\ref{#1}}
\newcommand{\Secs}[2]{Sections~\ref{#1} and \ref{#2}}
\newcommand{\Tab}[1]{Table~\ref{#1}}

\newcommand{\mps}{m~s$^{-1}$}
\newcommand{\cmss}{cm$^2$~s$^{-1}$}

\begin{document}

\title{Super-criticality of dynamo limits the memory of polar field to one cycle}

\email{karak.phy@iitbhu.ac.in}

\author{Pawan Kumar, Bidya Binay Karak, and Vindya Vashishth}




\affiliation{Department of Physics, Indian Institute of Technology (Banaras Hindu University), Varanasi, India 221005}







\begin{abstract}

The polar magnetic field precursor is considered to be the most robust and physics-based method for the prediction of the next solar cycle strength. However, to make a reliable prediction of a cycle, is the polar field at the solar minimum of the previous cycle enough or we need the polar field of many previous cycles? To answer this question, we performed several simulations using Babcock--Leighton type flux transport dynamo models with stochastically forced source for the poloidal field ($\alpha$ term). We show that when the dynamo is operating near the critical dynamo transition or only weakly supercritical, the polar field of cycle $n$
determines the amplitude of the next several cycles (at least three). However, when the dynamo is substantially supercritical, this
correlation of the polar field is reduced to one cycle. This change in the memory of the polar field from multi- to one-cycle 
with the increase of the super-criticality of the dynamo is independent of the importance of various turbulent transport processes in the model.
We further show that when the dynamo operates near the critical, 
it produces frequent extended episodes of weaker activity, resembling the solar grand minima. Occurrence of grand minima is accompanied with the multi-cycle correlation of polar field. The frequency of grand minima decreases with the increase of supercriticality of the dynamo. 
\end{abstract}



\section{Introduction} \label{sec:intro}
The large-scale magnetic field of the Sun oscillates in varying amplitude.
This field is believed to be generated through a large-scale dynamo, operating inside the convection zone (CZ) \citep{Mof78, C14, Char20}.
This dynamo is driven by the helical convection and the differential rotation.
In the $\alpha \Omega$ dynamo scenario (which is applicable for the sun),
the magnetic field grows when the dynamo number
\begin{equation}
 D = \frac{\alpha_0 \Delta \Omega R_\odot^3}{\eta_0^2}
\label{eq1}
\end{equation}
exceeds a critical value, where $\eta_0$ is the turbulent magnetic diffusivity,
$\Delta \Omega $ is angular velocity variation, and $\alpha_0$
is the measure of the $\alpha$ effect \citep{KR80}.
While, in the original $\alpha \Omega$ dynamo model, the helical convective $\alpha$
generates poloidal field from the toroidal one, in recent years, it has been realized
that the poloidal field in the sun is primarily generated through the decay of
tilted bipolar magnetic regions (BMRs), so-called the Babcock--Leighton process \citep{Ba61,Le64,Das10,KO11,Muno13,Priy14}.

With increasing age, stars spin down through the magnetic braking
\citep{Skumanich72},
thereby decreasing
the ability to generate the magnetic field. Thus there must be a critical age beyond which
each star ceased to produce a large-scale magnetic field. Stellar observations indeed find
pieces of evidence of a violation of the rotation-age relation of some stars \citep{RG15}. Further,
there is an upper value in the rotation period for each spectral type \citep{R84}, also see \citet{Met16}.
These observations can be explained by the ceased of the large-scale dynamo above a
certain rotation period; see discussion in \citet{KN17, CS17}.
Thus, the solar dynamo is possibly operating not too far from the critical
to the dynamo transition. Three-dimensional numerical simulations \citep{KKB15} 
and mean-field 
models \citep{KO10} showed that a slightly subcritical dynamo is also possible when the initial condition for the magnetic field is strong. 
On the other hand, explaining the deviation of gyrochronology in dynamo model, 
\citet{KN17} predicted that solar dynamo is about $10\%$ supercritical.

Besides understanding the physics of dynamo action, the prediction of the solar magnetic cycle
has become increasingly important due to its effect on the space and Earth's climate \citep{Petrovay20}.
It has been realized that in the \bl\ dynamo, the prediction is possible if we have knowledge
of the polar field at the preceding minimum \citep{Sch78, CCJ07}.
It is the polar field that is transported to the deep CZ where shear induces a toroidal
field for the following cycle \citep{CD00,JCC07, CB11}. \citet{YNM08} showed that when diffusive
transport dominates over the advective transport via meridional flow,
the memory of the polar field is limited to one cycle. 
However, when the advective transport dominates, the memory
can last for multiple cycles. 
Limited observations of polar faculae hints one cycle memory \citep{Muno13}. 
If this result is true in the Sun, then this means that to make a reliable prediction 
we must use a diffusion-dominated dynamo model in which we need the knowledge of the polar field of the previous one cycle.

The solar dynamo is nonlinear. As the differential rotation does not change with the
solar cycle (except a tiny variation in the form of torsional oscillation),
we expect the $\Omega$ effect i.e., the poloidal $\rightarrow$ toroidal, is not heavily nonlinear.
Some global convection simulations, however, show a considerable variation of differential rotation \citep[e.g.,][]{Kar15, Kap16}.
In contrast, the flux emergence and the \bl\ process i.e., toroidal $\rightarrow$ poloidal part, 
could be nonlinear. BMR tilt quenching \citep{KM17, LC17, Jha20}, 
active region inflow \citep{MC17}, and the latitudinal variation of BMR, the so-called latitudinal quenching \citep{J20,Kar20}
are the possible candidates
for the nonlinearity in the latter process.
As the flux emergence, BMR tilt and meridional flow involve some variation, the 
toroidal $\rightarrow$ poloidal part also involves some randomness.
The randomness in this part tries to reduce the memory of the polar flux.
In this study, we explore the importance of nonlinearity in the toroidal $\rightarrow$ poloidal part
on the memory of the polar flux.
Using stochastically forced kinematic \bl\ dynamo models we show that 
in addition to the stochasticity, it is the nonlinearity in toroidal $\rightarrow$ poloidal part
that determines the memory of the polar flux. 
When the dynamo is operating in the highly supercritical regime, 
the correlation between the toroidal to the polar flux of the same cycle 
is lost and then the multiple-cycle memory of the polar flux is limited to one cycle only.
This one cycle memory is independent of the relative importance of diffusion or advection processes which is a clear contradiction to \citet{YNM08}.

\section{Models}
\label{sec:mod}
We use two types of \bl\ dynamo models, namely, the flux transport and time delay dynamo models.
\subsection{Flux Transport Dynamos}
\label{sec:ftdm}
In the flux transport dynamo model \citep{Cha10, Kar14a}, we solve  following equations for axisymmetric 
magnetic fields.
\begin{equation}
 \frac{\partial A}{\partial t} + \frac{1}{s}({\bf v}\cdot{\bf \nabla})(s A)
 = \eta_{p} \left( \nabla^2 - \frac{1}{s^2} \right) A + S_\alpha,~~~~~~~~~~~~~~~~~~~~~~~~~~~~~~~~~~~~~~~~
\label{eq:pol}
\end{equation}
\begin{eqnarray}
 \frac{\partial B}{\partial t} 
 + \frac{1}{r} \left[ \frac{\partial}{\partial r}
 (r v_r B) + \frac{\partial}{\partial \theta}(v_{\theta} B) \right]
 = \eta_{t} \left( \nabla^2 - \frac{1}{s^2} \right)B ~~~~~~~~~~~~~~ \nonumber\\
 + s({\bf B}_p.\nabla)\Omega
 + \frac{1}{r}\frac{d\eta_t}{dr}\frac{\partial{B}}{\partial{r}},~~~~~~~~~~~~~~
\label{eq:tor}
\end{eqnarray}
where $A$ and $B$ are the potential of the poloidal magnetic field ($\bf {B}_p$) and
the toroidal magnetic field, respectively such that
${\bf B} =  {\bf B}_p + B {\bf \hat{e}_{\phi}}$,
with ${\bf B}_p = {\bf \nabla} \times A {\bf \hat{e}_{\phi}}$,
$s = r \sin \theta$ with $\theta$ being the colatitude,
${\bf v}=v_r {\bf \hat{e}_r} + v_{\theta} {\bf \hat{e}_{\theta}}$
is the meridional circulation, $\Omega$ is the angular velocity, $\eta_p$ and $\eta_t$ are the diffusivities of the
poloidal and toroidal fields, respectively, $S_\alpha$ is the parameter which captures the \bl\ process for the generation of poloidal field from toroidal one. 
We use six models, namely Models~I, II, III, IV, V and VI. 

\subsubsection{Models I--II}
\label{sec:modsI-II}
For Models~I and II, we use 
the local prescription of 
$\alpha$, which was done in the Surya code \citep{NC02, CNC04, C18}. 
In these models, $S_\alpha = \alpha B$, where
\begin{eqnarray}
\alpha = \frac{\alpha_0}{4}\cos\theta \left[1 + \mathrm{erf} \left(\frac{r - 0.95R_\odot}{0.025R_\odot}\right) \right]\nonumber \\
\times \left[1 - \mathrm{erf} \left(\frac{r - R_\odot}{0.025R_\odot}\right) \right].
\label{eq:alpha}
\end{eqnarray}
For the present study,
we use the same model as given in \citet{YNM08}. For Model~I, we have 
used the parameters 
for the poloidal field diffusion: $\eta_2 = 1\times10^{12}$~\cmss\ and $\eta_0 = 2\times10^{12}$~\cmss\ and for the meridional circulation: $v_0 =$ 15~\mps, 
while for Model~II, we use the same diffusivities, but $v_0 =$ 26~\mps.
\citet{YNM08} call these two models as the diffusion-dominated regime (their Run~1)
and advection-dominated regime (Run~2; see their Section 5.1).

\subsubsection{Models III--VI}
\label{sec:modsIII-IV}
For Models III--VI, we use the nonlocal $\alpha$ prescription, as initially done in \citet{DC99}, 
and followed by many authors; see \citet{CNC05,CH16} for discussion on different $\alpha$ prescriptions. 
Thus in this model,
\begin{equation}
 S_\alpha = \frac{\alpha}{1+\left(\frac{B(0.7R_\odot,\theta)}{B_0}\right)^2} B(0.7R_\odot,\theta),
\label{eq:alphaIII}
\end{equation}
where
\begin{eqnarray}
\alpha=\frac{\alpha_0}{4}
\sin\theta\cos\theta\left[\frac{1}{1+e^{-\gamma(\theta-\pi/4)}}\right] \nonumber \\
\left[1+\mathrm{erf}\left(\frac{r-0.95R_\odot}{0.05R_\odot}\right)\right]\left[1-\mathrm{erf}\left(\frac{r-R_\odot}{0.01R_\odot}\right)\right]
\label{alphaprof_nonlocal}
\end{eqnarray}
with $\gamma=30$ and $B_0 = 4 \times 10^4$~G.
Also, $\eta_t=\eta_p = \eta$, where
\newcommand{\etaRZ}{\eta_{\mathrm{RZ}}}
\newcommand{\etasurf}{\eta_{\mathrm{surf}}}
\def\Rs{R_{\odot}}
\begin{eqnarray*}
\eta(r) = \etaRZ + \frac{\eta_0}{2}\left[1 + \mathrm{erf} \left(\frac{r - 0.7\Rs}
{0.02\Rs}\right) \right] \\ \nonumber
+\frac{\etasurf}{2}\left[1 + \mathrm{erf} \left(\frac{r - 0.9\Rs}
{0.02\Rs}\right) \right],
\label{eq:eta}
\end{eqnarray*}
with 
$\etaRZ = 5\times10^{8}$~\cmss\ and
$\etasurf =  2\times10^{12}$~\cmss.
Basically, we use the same model as
given in the Reference Solution of \citet{HY10a} 
with only change in the value of $\eta_{\rm surf}$. 
For Models~III, V, and VI, we use $\eta_0 =  5\times10^{10}$~\cmss,
while for Model~IV, we use five times less diffusivity than in Model~III. 
Model~V is the same as Model~III but the meridional circulation is switched off.

\subsection{Time Delay Dynamo}
\label{sec:moddelay}
Finally, we use a time delay dynamo model which was developed in \citet{wilsmith}; also see \citet{Ha14} for an application of this model. 
In this model,
the equations for the toroidal and poloidal fields are truncated by removing the spatial dependences 
and some time delays are introduced to mimic the finite times required to communicate the fields between the BCZ to the surface 
through meridional flow and magnetic buoyancy. Following equations are solved in this model.

\begin{equation}
\frac{\mathrm {d} B}{\mathrm{ d} t} = \frac{\omega}{L} A (t - T_0) - \frac{B}{\tau_d}
\end{equation}
\begin{equation}
\frac{\mathrm {d} A}{\mathrm{ d} t} = \frac{\alpha_0} { 1 + \left[  \frac{ B (t - T_1) }{ B_{\rm eq} } \right]^2 } B (t - T_1) - \frac{B}{\tau_d}.
\label{eq:delay2}
\end{equation}
Here, $T_0$ represents the time delay required for the generation of toroidal field from the poloidal one
through differential rotation, while $T_1$ is the time delay involved in the production of 
poloidal field from the toroidal field through \bl\ process. $\omega$ and $L$ are the contrast in 
differential rotation and the length scale in the tachocline, respectively, $\tau_d$ is the diffusion time scale of the
turbulent diffusion in the CZ, $\alpha_0$ is the amplitude of the $\alpha$ (like the one in flux transport dynamo models). We note that unlike in the previous delay dynamo model of \citet{wilsmith}, here we have included the usual alpha quenching: $1 / (1 + (B/B_{\rm eq})^2)$.
The parameters for our study are as follows: 
$T_0 = 2$, $T_1= 0.5$, $w/L= - 0.34$, $B_{\rm eq} = 1$, $\alpha_0 = 0.29$, and $\tau_d = 15$.

\begin{figure}
\centering
\includegraphics[scale=0.35]{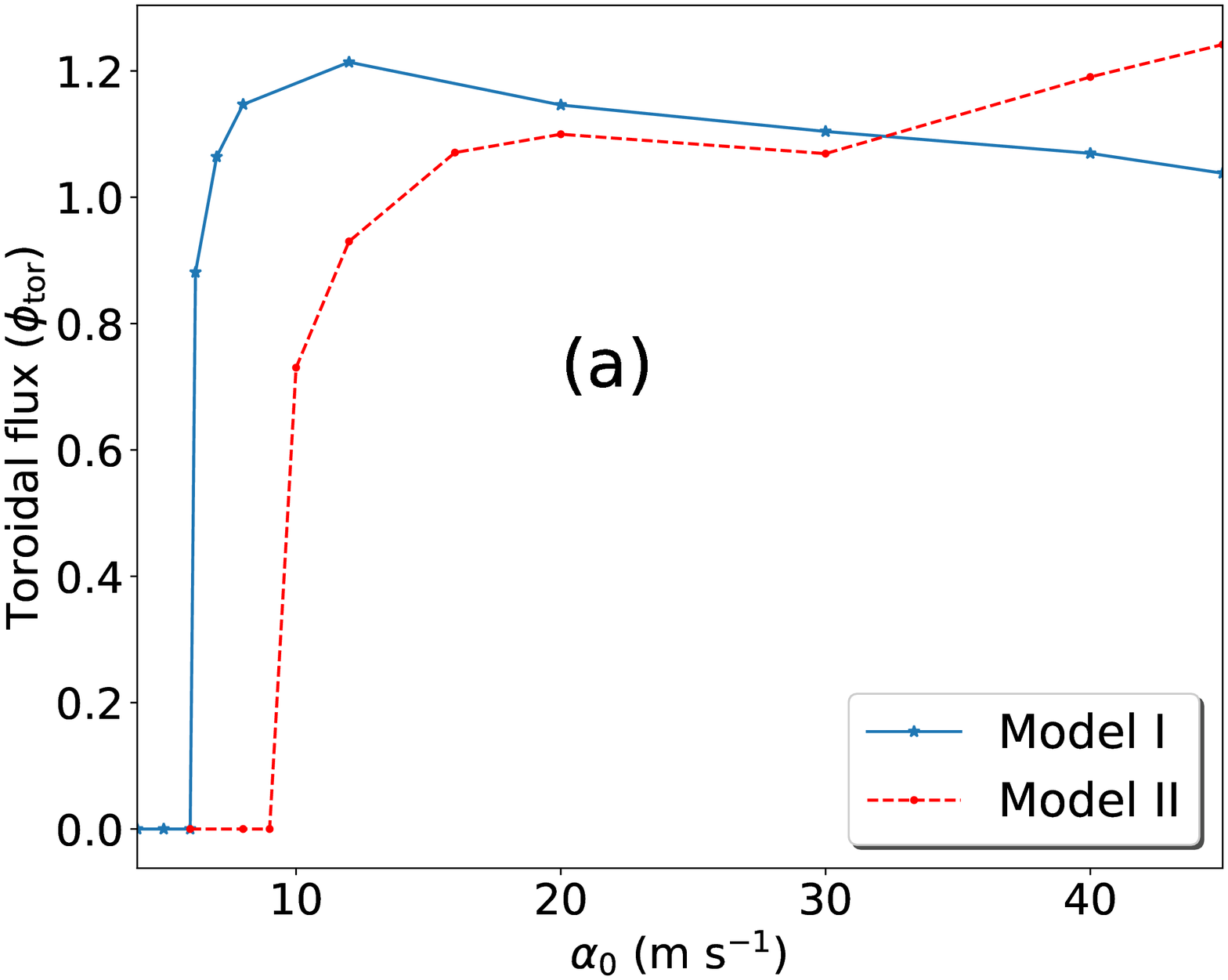}
\includegraphics[scale=0.35]{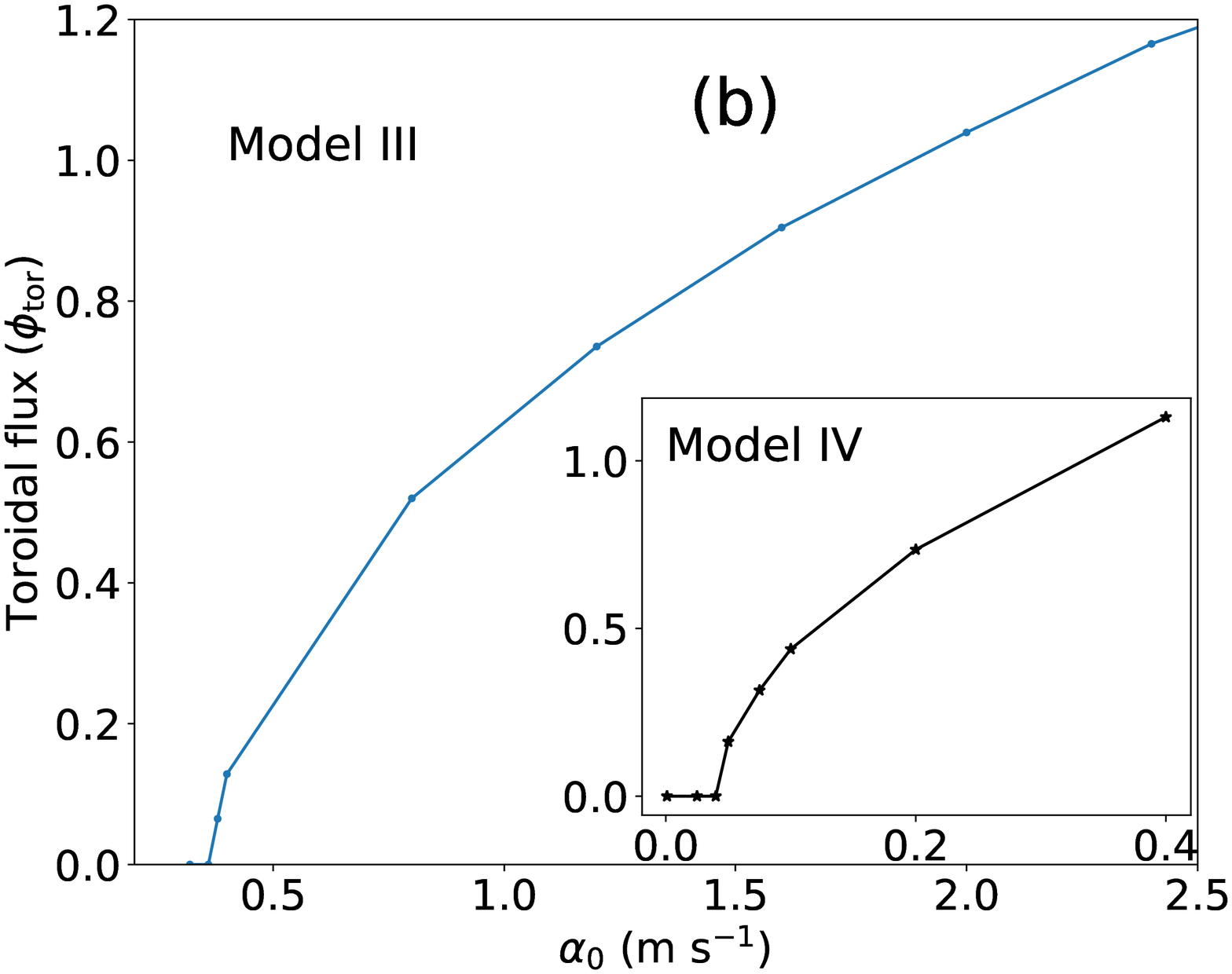}
\includegraphics[scale=0.35]{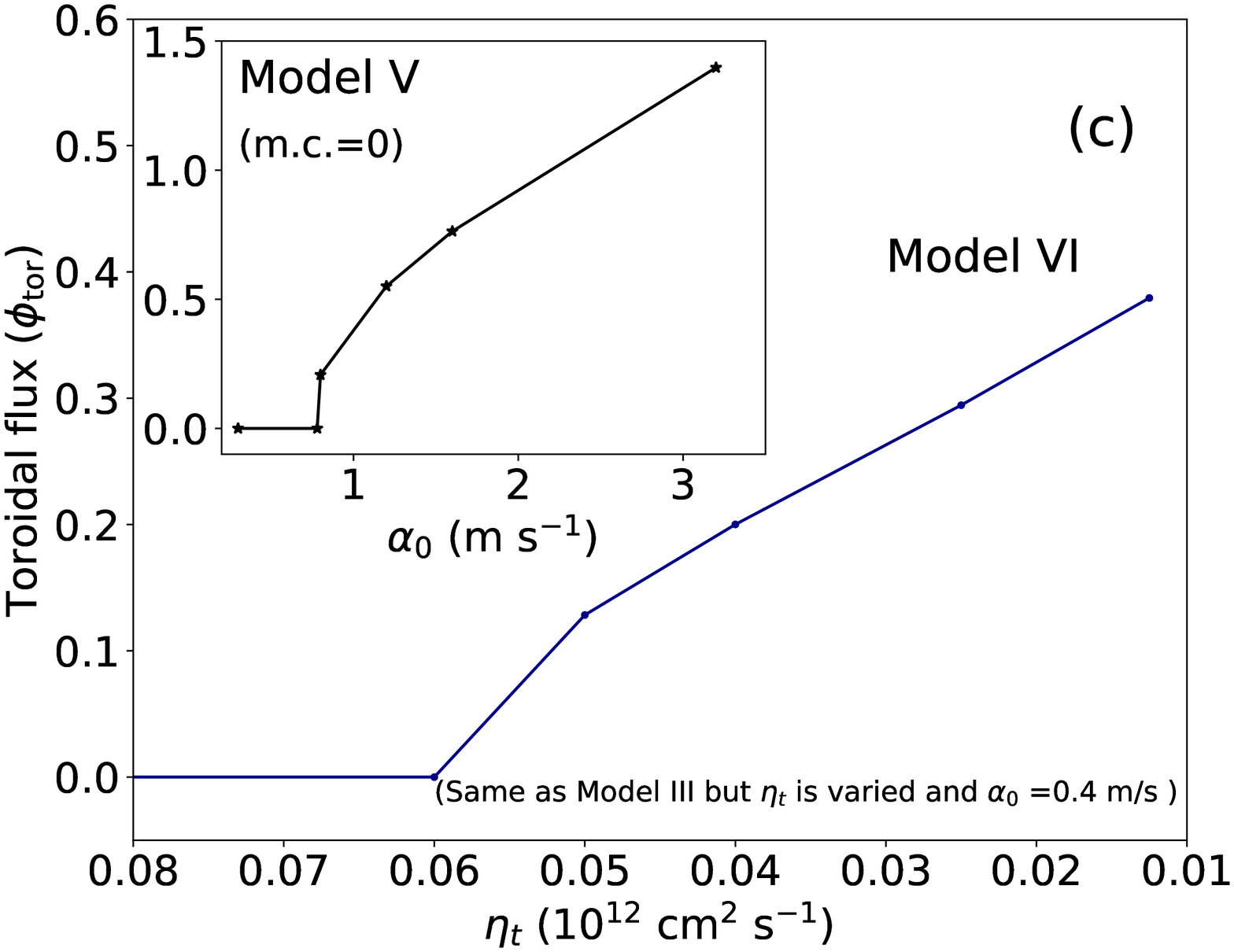}
\caption{
(a): Variation of the toroidal flux $\phi_{\rm tor}$ with the strength 
of the Babcock-Leighton $\alpha_0$ from Models~I (solid) and II (dashed).
 The critical $\alpha_0$, $\alpha_0^{\rm crit}$ for Models I and II are 6.2~\mps\ and 10~\mps.
(b): Same as (a) but obtained from Model~III and IV (shown in inset);
note in these models, $\alpha_0^{\rm crit} = 0.38$ and 0.05~\mps.
(c): For Model~VI 
($\eta_0$ is varied; note $\eta_0^{\rm crit} = 0.05\times10^{12}$~\cmss).
(c) Inset: for Model V 
(no meridional circulation); 
$\alpha_0^{\rm crit} =$ 0.8~\mps.
}
\label{fig:dynamo}
\end{figure}

\section{Results from Flux Transport Dynamo Models}
\label{sec:res1}
We shall first present the results of our flux transport dynamo models in details. 
\subsection{Identifying critical dynamo parameters}
Let us identify the dynamo transitions in each model. 
As seen in \Eq{eq1}, the growth of the magnetic field is possible when $D$ exceeds
a critical value. In our models, this is possible either by increasing $\alpha_0$ or by
decreasing $\eta_0$. 
In Models~I--V, 
we vary $\alpha_0$  while keeping other parameters same. \Fig{fig:dynamo} shows
the mean $\phi_{\rm tor}$ from these runs; $\phi_{\rm tor}$ is computed over a
layer $r = 0.677 R_\odot$ -- $0.726 R_\odot$ and latitudinal extent $10^\circ$--$45^\circ$,
and normalized by the area and $B_0$.
In Model~VI, we fix $\alpha_0$ at $0.4$~\mps\ and decrease $\eta_0$; \Fig{fig:dynamo}(c) shows
this result. We find that the critical $\alpha_0$, defined as $\alpha_0^{\rm crit}$, 
for the dynamo transition in Models~I, II, III, IV, and V are 6.2, 10, 0.38, 0.05 and 0.8~\mps.  
In Model~VI, where the critical $\eta_0$, defined as $\eta_0^{\rm crit}$, is around $0.05 \times 10^{12}$~\cmss.
As long as $\alpha_0$ is above these critical values (or $\eta_0$ below $\eta_0^{\rm crit}$), the magnetic field 
shows a regular polarity reversal. Increasing $\alpha_0$ (or decreasing $\eta_0$) makes the model more supercritical.
In the supercritical regime, the magnetic field does not grow linearly because of the nonlinearity
imposed in these kinematic dynamo models.

Let us make a few comments on \Fig{fig:dynamo}. We observe that in Models III--V, 
the variation of magnetic field with $\alpha_0$ for $\alpha_0 > \alpha_0^{\rm crit}$ 
is different than that in Models I and II. This difference is because of different ways
of including nonlinearity in these models. In Models I and II (Surya/local $\alpha$), the nonlinearity is included in 
the magnetic buoyancy part in the following way. When the toroidal field in any latitude grid
above the base of the CZ at intervals of time $8.8\times10^5$~s exceeds a
certain value ($B_0 = 0.8\times10^5$~G), $50\%$ of the flux at that grid is reduced 
and the same is deposited at a surface layer. This reduction of toroidal flux 
is the only nonlinearity in Models I and II. However, in Models III--VI, 
a nonlinearity of the form $1/[1+(B/B_0)^2]$ is 
included in \bl\ $\alpha$ term; see \Eq{eq:alphaIII}.

\subsection{Identifying the correlation of polar field}
\label{sec:rescorr}
Now we include stochastic fluctuations in the \bl\ $\alpha$ to produce variable 
magnetic cycles at different regimes of the dynamo i.e., at different dynamo parameters.
To do so, in \Eq{eq:alpha} and \Eq{alphaprof_nonlocal} we replace $\alpha_0$ by
$\alpha_0 (1 + \sigma (t,\tau_{\rm cor}))$, where
$\sigma$ is uniform random deviate within $[-1, 1]$, and $\tau_{\rm cor}$ is the
time after which $\alpha_0$ is updated. 
The value of  $\tau_{\rm cor}$ is chosen in such a way that the ratio of cycle period 
to $\tau_{\rm cor}$  remains the same. For reference, $\tau_{\rm cor} = 2.3$~yr and 1.5~yr 
respectively for Models~I and II when $\alpha_0 = 30$~\mps.
Unless stated otherwise, in all stochastically forced simulations, we include $100\%$ 
fluctuations\footnote{\citet{YNM08} called it $200\%$ although they used the same level of fluctuations (private communication). 
We believe that it is just the convention of measurement of the percentage of fluctuation level.}.
Thus, the relative fluctuation level is kept the same.
We perform simulations at different values of 
$\hat{\alpha}_0 = \alpha_0/\alpha_0^{\rm crit})$ (for Models I-V)
and $\hat{\eta}_0 = \eta_0^{\rm crit}/\eta_0$ (for Model~VI). 

We realized when $100\%$ fluctuations in $\alpha_0$ is added, Models I and II (Surya) tends to decay
unless $\alpha_0$ is sufficiently above $\alpha_0^{\rm crit}$. Or in other words, the $\alpha_0^{\rm crit}$
is increased in Models I and II when fluctuations are included. 
This does not happen in Models III--V (nonlocal $\alpha$).
This different behavior is due to extensively different parameters and the treatment of magnetic buoyancy 
in these models. 
Therefore, with $100\%$ fluctuations in $\alpha_0$ we obtain stable cycles in Models I and II only 
when $\hat{\alpha}_0 \ge 2.0$.
In \Tab{table1},
we present the correlations between the peaks of the polar flux and 
the peaks of the low-latitude toroidal
flux at the base of CZ ($\phi_{\rm r}$ is computed on the solar surface over the latitudinal extent $70^\circ$--$89^\circ$,
and normalized by the area and $B_0$).

\begin{figure}
\centering
\includegraphics[scale=0.45]{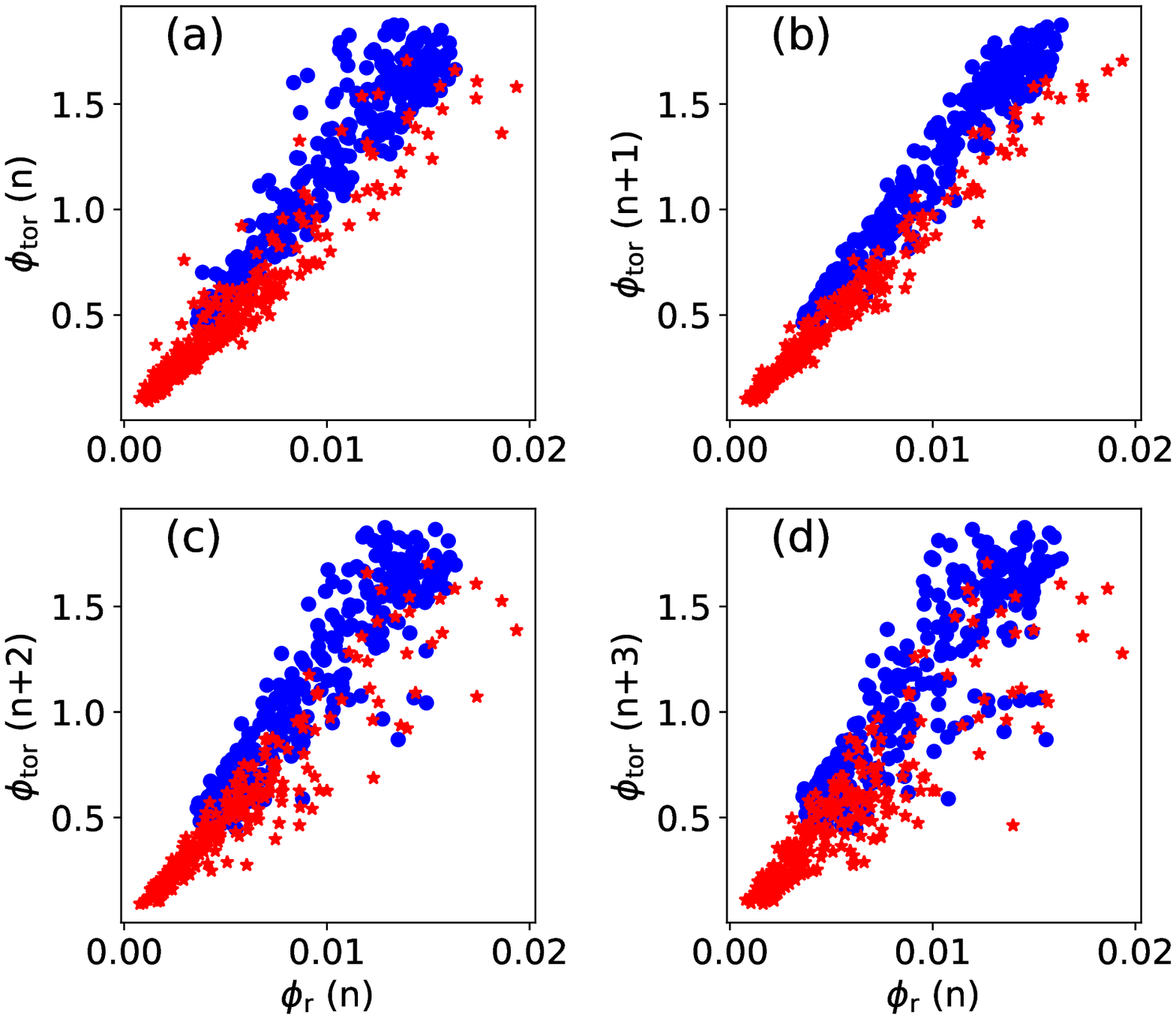}
\includegraphics[scale=0.45]{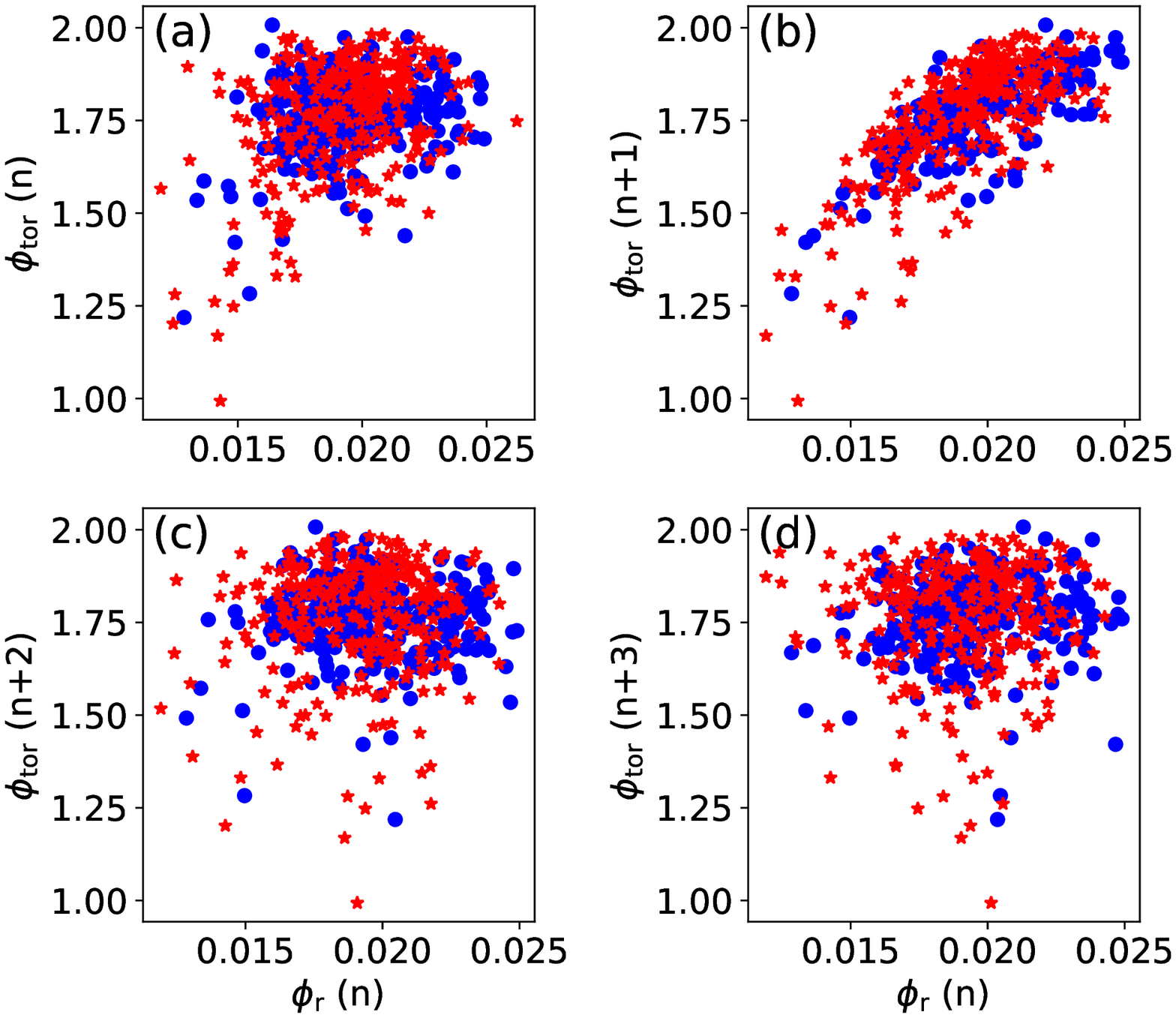}
\caption{
Scatter plots between the peaks of the surface polar flux of cycle $n$, $\phi_r (n)$ 
with that of the low-latitude toroidal flux at the base of CZ, $\phi_{tor}$ of cycle 
(a) $n$, (b) $n+1$, (c) $n+2$, and (d) $n+3$ from Models~I (blue/circles) and II (red/asterisks). 
Top four and bottom four panels are obtained from runs at weakly supercritical 
($\hat{\alpha}_0 = 2$) and highly supercritical ($\hat{\alpha}_0 = 4$) dynamos, respectively.
For Model~I data, a factor of two is multiplied to $\phi_r (n)$ in top four panels
and in the bottom four panels 2.8 and 1.04 are multiplied to $\phi_r (n)$ and $\phi_{\rm tor} (n)$, respectively for comparison.
}
\label{fig:dicorr}
\end{figure}

We first consider the results from Models I and II which are
shown on the left columns of \Tab{table1} and \Fig{fig:dicorr}. 
We observe that when $\hat{\alpha}_0$ 
is small, $\phi_{\rm r} (n)$ correlates strongly with $\phi_{\rm tor}$ of cycle $n$, $n+1$, $n+2$ and $n+3$.
These multiple-cycle correlations decreases with the increase 
of $\hat{\alpha}_0$.
In the highly supercritical regime (large $\hat{\alpha}_0$), 
$\phi_{\rm r} (n)$ strongly correlates with $\phi_{\rm tor}$ of cycle
$n+1$ alone.


%

\begin{table*}
\centering
\caption{Correlation coefficients between $\phi_r$ (n) and $\phi_{\rm tor}$ of different cycles 
for simulations at different values of $\hat{\alpha}_0$ and $\hat{\eta}_0$.}
\begin{tabular}{clcccclccccclc} 

\cline{1-14}

\multicolumn{4}{c}{Local $\alpha$ (Surya)} && \multicolumn{5}{c}{Nonlocal $\alpha$}&& \multicolumn{3}{c}{Nonlocal $\alpha$}\\
\cline{1-4}
\cline{6-10}
\cline{12-14}
 & & Model I & Model II && && Model III & Model IV & Model V && &  Model~VI\\
\cline{1-4}
\cline{6-10}
\cline{12-14}
$\hat{\alpha}_0$ & $\phi_r$ (n) \&  & $r$ (s.l.?) & $r$ (s.l.?) && $\hat{\alpha}_0$ & $\phi_r$ (n) \&  & $r$ (s.l.?) & $r$ (s.l.?) & $r$ (s.l.?) && $\hat{\eta}_0$ & $\phi_r$ (n) \&  & $r$ (s.l.?)\\
\cline{1-4}
\cline{6-10}
\cline{12-14}
 2.0& $\phi_{\rm tor}$ (n)  & 0.94 (Y) & 0.96 (Y)&& 1.0 & $\phi_{\rm tor}$ (n)  & 0.92 (Y)& 0.74 (Y)& 0.86 (Y) && 1.0 & $\phi_{\rm tor}$ (n)  & 0.79 (Y)\\
    & $\phi_{\rm tor}$ (n+1)& 0.96 (Y) & 0.98 (Y)&&     & $\phi_{\rm tor}$ (n+1)& 0.99 (Y)& 0.99 (Y)& 0.94 (Y)&&     & $\phi_{\rm tor}$ (n+1)  & 0.99 (Y)\\
    & $\phi_{\rm tor}$ (n+2)& 0.90 (Y) & 0.96 (Y)&&     & $\phi_{\rm tor}$ (n+2)& 0.91 (Y)& 0.78 (Y)& 0.77 (Y)&&     & $\phi_{\rm tor}$ (n+2)  & 0.77 (Y)\\
    & $\phi_{\rm tor}$ (n+3)& 0.86 (Y) & 0.92 (Y)&&     & $\phi_{\rm tor}$ (n+3)& 0.89 (Y)& 0.68 (Y)& 0.60 (Y)&&     & $\phi_{\rm tor}$ (n+3)  & 0.73 (Y)\\
\cline{1-4}
\cline{6-10}
\cline{12-14}
 2.5& $\phi_{\rm tor}$ (n)  & 0.71 (Y) & 0.69 (Y)&& 1.5 & $\phi_{\rm tor}$ (n)  & 0.23 (Y)& 0.39 (Y)& 0.32 (Y)&& 1.5 & $\phi_{\rm tor}$ (n)  & 0.50 (Y)\\
    & $\phi_{\rm tor}$ (n+1)& 0.80 (Y) & 0.87 (Y)&&     & $\phi_{\rm tor}$ (n+1)& 0.99 (Y)& 0.97 (Y)& 0.89 (Y)&&     & $\phi_{\rm tor}$ (n+1)  & 0.99 (Y)\\
    & $\phi_{\rm tor}$ (n+2)& 0.54 (Y) & 0.79 (Y)&&     & $\phi_{\rm tor}$ (n+2)& 0.23 (Y)& 0.47 (Y)& 0.35 (Y)&&     & $\phi_{\rm tor}$ (n+2)  & 0.48 (Y)\\
    & $\phi_{\rm tor}$ (n+3)& 0.47 (Y) & 0.59 (Y)&&     & $\phi_{\rm tor}$ (n+3)& 0.26 (Y)& 0.42 (Y)& 0.20 (Y)&&     & $\phi_{\rm tor}$ (n+3)  & 0.49 (Y)\\
\cline{1-4}
\cline{6-10}
\cline{12-14}
 3.0& $\phi_{\rm tor}$ (n)  & 0.40 (Y) & 0.59 (Y)&& 2.0 & $\phi_{\rm tor}$ (n)  &$-0.07$ (N)& 0.19 (Y)& 0.18 (Y)&& 2.0 & $\phi_{\rm tor}$ (n)  & 0.28 (Y)\\
    & $\phi_{\rm tor}$ (n+1)& 0.72 (Y) & 0.82 (Y)&&     & $\phi_{\rm tor}$ (n+1)& 0.99 (Y)  & 0.97 (Y)& 0.89 (Y)&&     & $\phi_{\rm tor}$ (n+1)  & 0.99 (Y)\\
    & $\phi_{\rm tor}$ (n+2)& 0.26 (Y) & 0.40 (Y)&&     & $\phi_{\rm tor}$ (n+2)&$-0.09$ (N)& 0.29 (Y)& 0.25 (Y)&&     & $\phi_{\rm tor}$ (n+2)  & 0.28 (Y)\\
    & $\phi_{\rm tor}$ (n+3)& 0.26 (Y) & 0.23 (Y)&&     & $\phi_{\rm tor}$ (n+3)& 0.18 (N)  & 0.20 (Y)& 0.07 (N)&&     & $\phi_{\rm tor}$ (n+3)  & 0.41 (Y)\\
\cline{1-4}
\cline{6-10}
\cline{12-14}
 4.0& $\phi_{\rm tor}$ (n)  & 0.27 (Y) & 0.37 (Y)&& 4.0 & $\phi_{\rm tor}$ (n)  &$-0.39$ (Y)&$-0.12$ (N)& 0.02 (N)&& 4.0 & $\phi_{\rm tor}$ (n)  & 0.03 (N)\\
    & $\phi_{\rm tor}$ (n+1)& 0.67 (Y) & 0.75 (Y)&&     & $\phi_{\rm tor}$ (n+1)& 0.99 (Y)  & 0.95 (Y)& 0.88 (Y)&&     & $\phi_{\rm tor}$ (n+1)  & 0.99 (Y)\\
    & $\phi_{\rm tor}$ (n+2)& $-0.01$ (N) & 0.10 (N)&&     & $\phi_{\rm tor}$ (n+2)&$-0.40$ (Y)& 0.00 (N)& 0.09 (N)&&     & $\phi_{\rm tor}$ (n+2)  & 0.04 (N)\\
    & $\phi_{\rm tor}$ (n+3)& 0.10 (N) &$-0.02$ (N)&&   & $\phi_{\rm tor}$ (n+3)& 0.25 (Y)  & 0.05 (N)& $-0.01$ (N)&&     & $\phi_{\rm tor}$ (n+3)  & 0.33 (Y)\\
\cline{1-4}
\cline{6-10}
\cline{12-14}
 5.0& $\phi_{\rm tor}$ (n)  & 0.15 (N) & 0.31 (Y)&&     &                       &            &           &&     &                     &            \\
    & $\phi_{\rm tor}$ (n+1)& 0.72 (Y) & 0.68 (Y)&&     &                       &            &           &&     &                     &             \\
    & $\phi_{\rm tor}$ (n+2)&$-0.03$ (N)& 0.03 (N)&&    &                       &            &           &&     &                     &            \\
    & $\phi_{\rm tor}$ (n+3)& 0.10 (N) & 0.04 (N)&&     &                       &            &           &&     &                     &            \\
\cline{1-14} 
\cline{1-14}
\end{tabular}
\tablecomments{Here $\hat{\alpha}_0 = \alpha_0/\alpha_0^{\rm crit}$ 
and $\hat{\eta}_0 = \eta_0^{\rm crit}/\eta_0$, where the superscript, `crit' 
represents the minimum (maximum) values of $\alpha_0$ ($\eta_0$) needed for 
the dynamo action in each model; see \Fig{fig:dynamo}. 
In Models I--V, $\alpha_0$ is increased, while in Model VI, $\eta_0$ is decreased in different runs.
275 data (cycles) are used to obtain the correlations. 
The `Y/N', the value of s.l., represents whether the correlation is statistically significant or not.
In all simulations,
$100\%$ fluctuations in $\alpha_0$ is included.
However, due to statistical uncertainties in numerical realizations, the correlation coefficients are slightly different 
when runs are repeated with different realizations of random numbers.
}
\label{table1}
\end{table*}

\begin{figure*}
\centering
\includegraphics[scale=0.3]{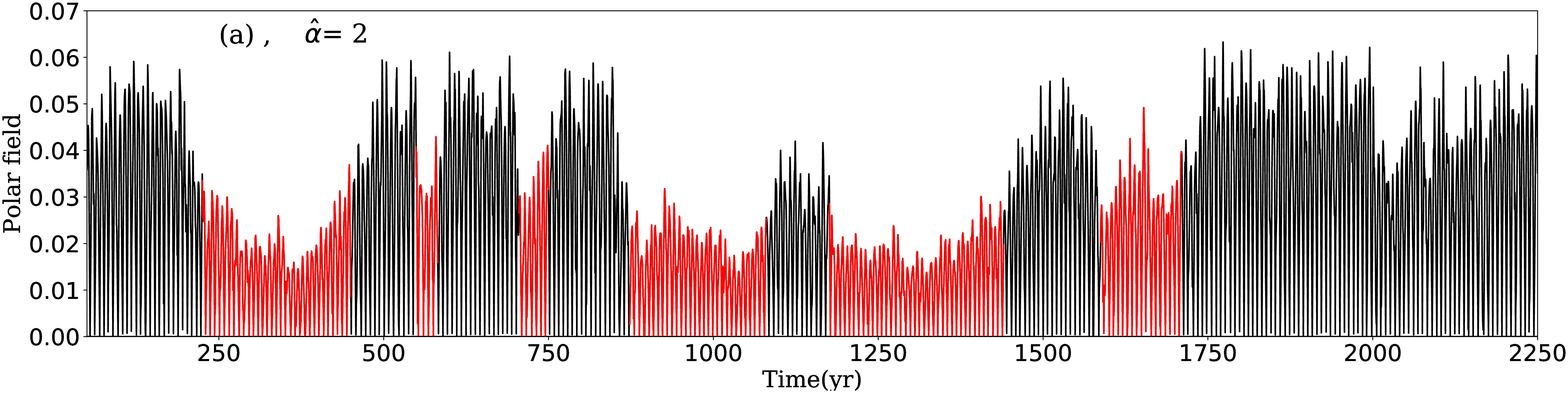}
\includegraphics[scale=0.3]{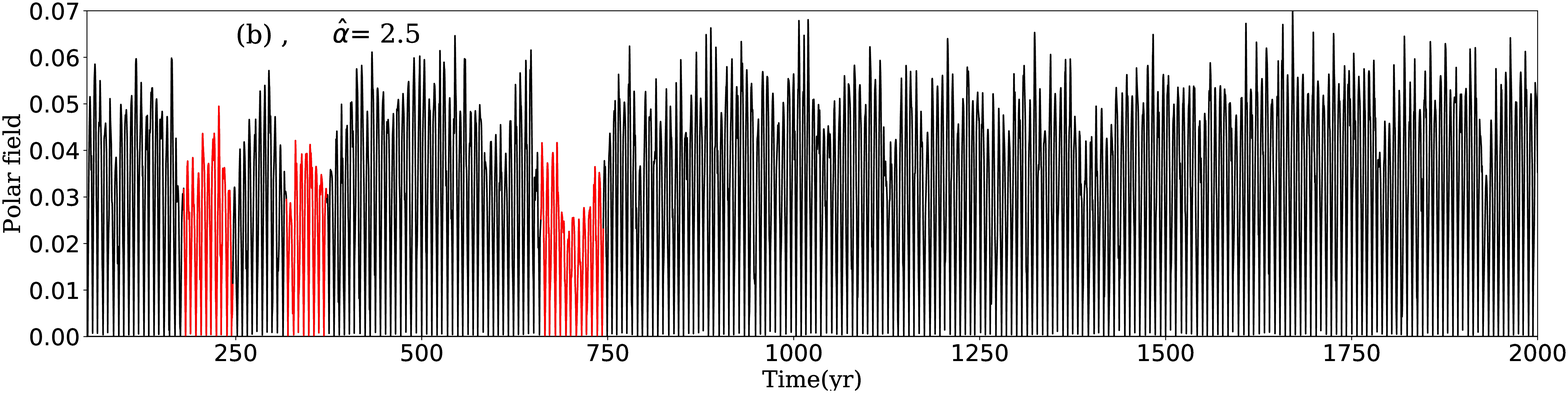}
\includegraphics[scale=0.3]{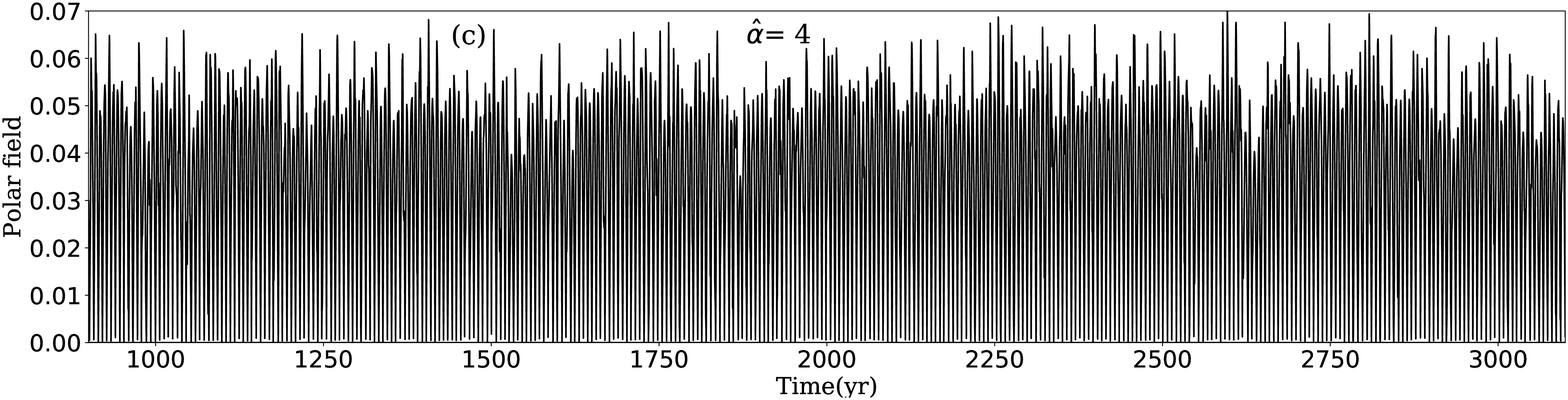}
\caption{
Time series of polar flux from Model I for (a) $\hat{\alpha} = 2$ (weakly supercritical), (b) $\hat{\alpha} = 2.5$, and (c) $\hat{\alpha} = 4$ (highly supercritical).
Red color highlights the weaker activity episodes.
}
\label{fig:ModelI}
\end{figure*}

\begin{figure}
\centering
\includegraphics[scale=0.45]{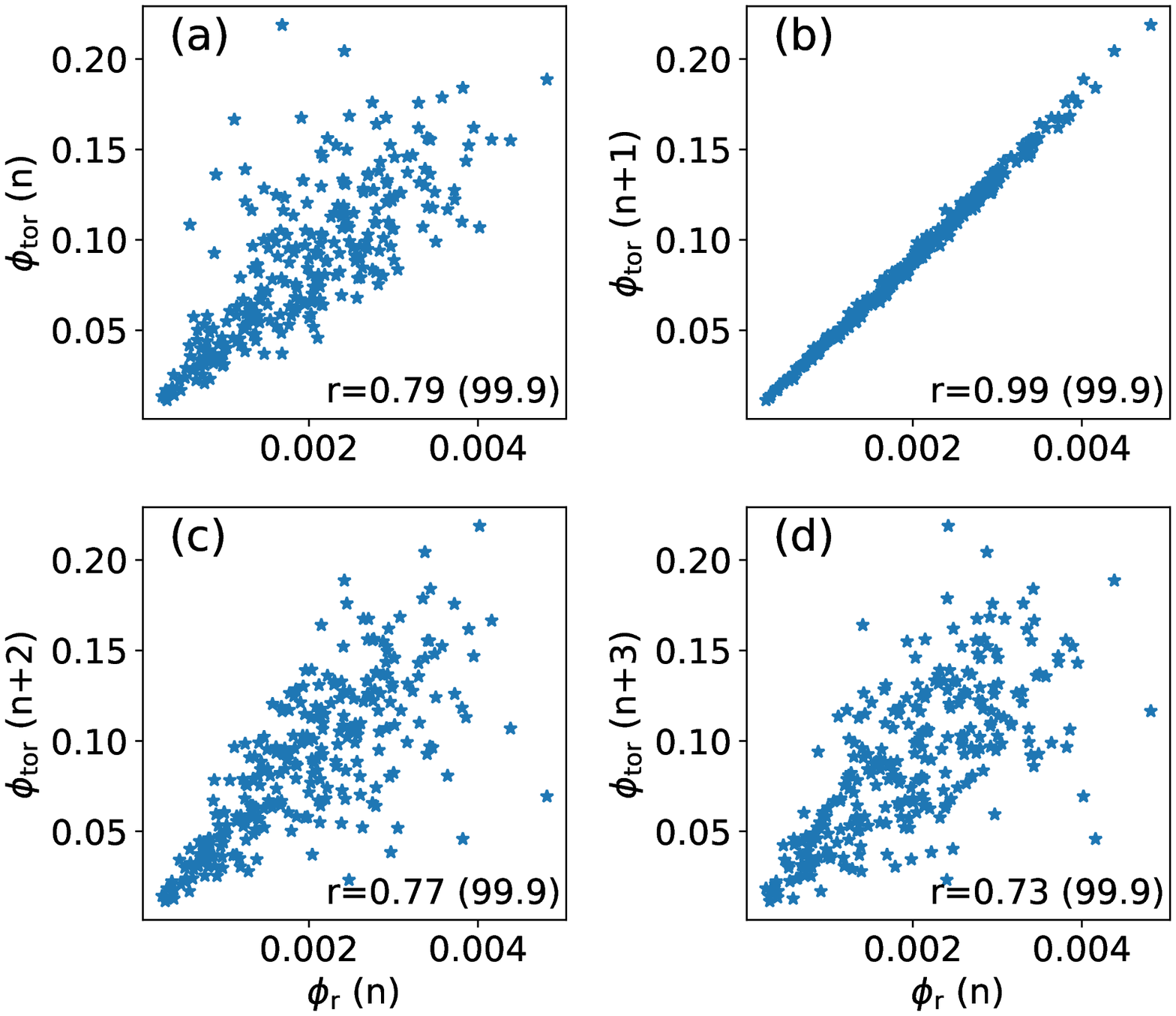}
\includegraphics[scale=0.45]{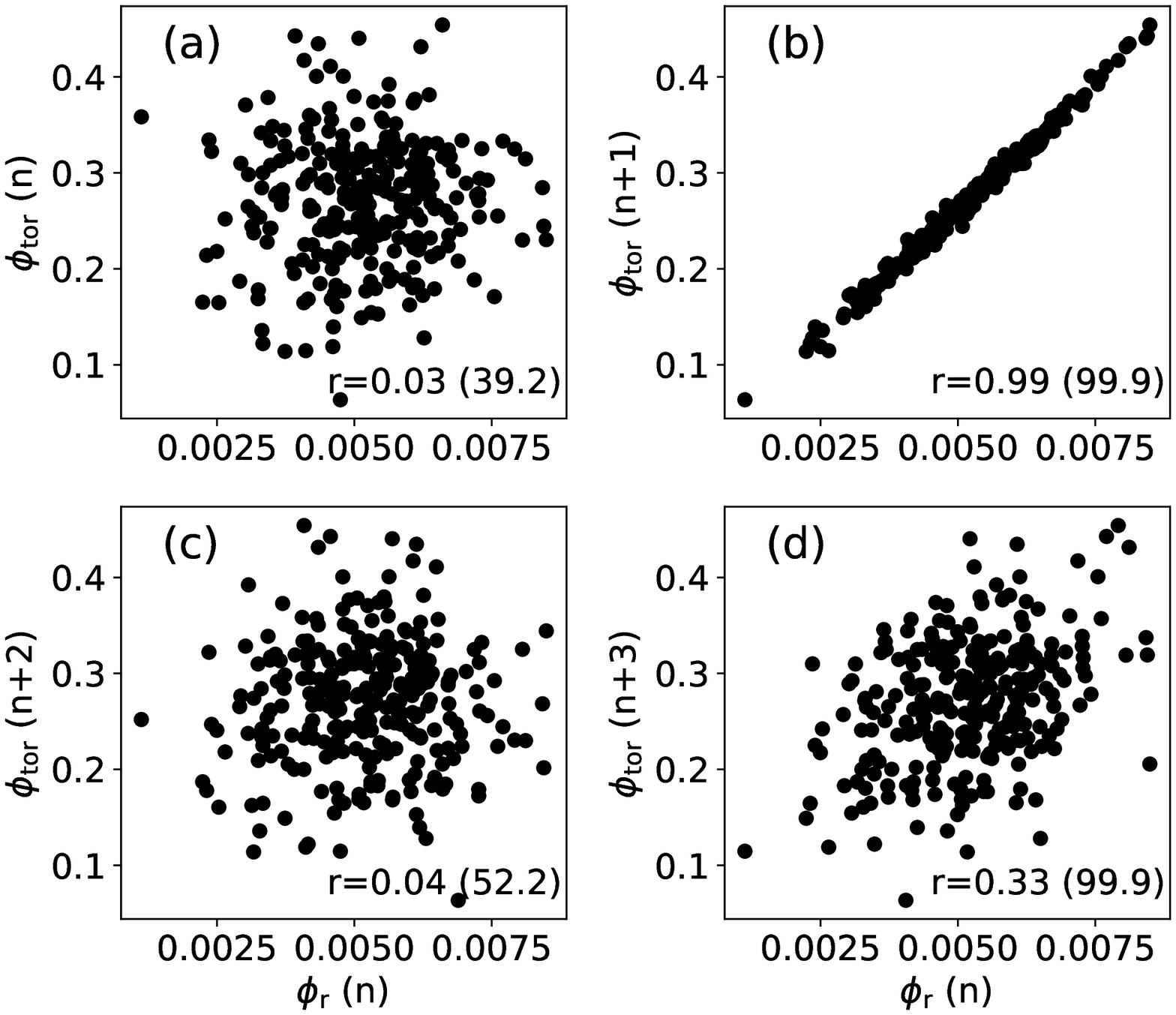}
\caption{
The format is same as \Fig{fig:dicorr} but obtained from Model~V at $\alpha_0=0.4$~\mps. 
Top four and bottom four panels 
are obtained from a weakly supercritical ($\hat{\eta}_0=1$) 
and highly supercritical dynamos ($\hat{\eta}_0=4$), respectively.
}
\label{fig:etavaried}
\end{figure}

%

\begin{table}
\centering
\caption{The mean cycle period $T_{\rm av}$ of the simulations presented in \Tab{table1}.
}
\begin{tabular}{clclcccclclcc}
\cline{1-11} 
 Model:&{I} &{II} && &{III} &{IV} & V &&\multicolumn{2}{c}{VI}\\
\cline{1-11}
&\multicolumn{2}{c}{(Local $\alpha$)} && &\multicolumn{3}{c}{(Nonlocal $\alpha$)} && \multicolumn{2}{c}{(Nonlocal $\alpha$)}\\
\cline{1-3}
\cline{5-8}
\cline{10-11}
$\hat{\alpha}_0$ & $T_{\rm av}$ & $T_{\rm av}$  & &$\hat{\alpha}_0$ & $T_{\rm av}$ & $T_{\rm av}$ & $T_{\rm av}$  & &$\hat{\eta}_0$ & $T_{\rm av}$ \\
\cline{1-3}
\cline{5-8}
\cline{10-11}
2.0 & 22.4 & 13.8 && 1.0 & 9.1& 9.3& 8.3 && 1.0 & 9.0\\
2.5 & 21.6 & 13.5 && 1.5 & 8.9& 9.1& 8.4 && 1.5 & 9.1\\
3.0 & 21.5 & 13.1 && 2.0 & 8.8& 8.9& 8.5 && 2.0 & 9.1\\
4.0 & 20.8 & 12.8 && 4.0 & 8.7& 8.6& 9.0 && 4.0 & 9.3\\
5.0 & 20.2 & 12.5 &&     &    &    & &&     &    \\
\cline{1-11} 
\end{tabular}
\label{table2}
\end{table}

\begin{table}
\centering

\caption{The same as Table 1 but only for Model IV and at increased fluctuation levels.}
\begin{tabular}{clclcc}
\cline{1-4}
\multicolumn{4}{c}{Nonlocal $\alpha$}\\
\cline{1-4}
 & & 150$\%$    & 200$\%$\\
 \cline{1-4}
$\hat{\alpha}_0$ & $\phi_r$ (n) \&  & $r$ (s.l.?)   & $r$ (s.l.?) \\

\cline{1-4}
1.0&$\phi_{\rm tor}$ (n)  & 0.80 (Y) & 0.83 (Y)\\ 
& $\phi_{\rm tor}$ (n+1) & 0.97 (Y) & 0.98 (Y)\\
    & $\phi_{\rm tor}$ (n+2)& 0.79 (Y) & 0.85 (Y)\\
    & $\phi_{\rm tor}$  (n+3)& 0.71 (Y) & 0.82 (Y)\\
\cline{1-4}
1.5 & $\phi_{\rm tor}$ (n)  & 0.40 (Y) & 0.35 (Y)\\ 
& $\phi_{\rm tor}$ (n+1) & 0.97 (Y) & 0.96 (Y)\\
    & $\phi_{\rm tor}$ (n+2)& 0.48 (Y) & 0.42 (Y)\\
    & $\phi_{\rm tor}$  (n+3)& 0.43 (Y) & 0.37 (Y)\\
\cline{1-4}
2.0 &$\phi_{\rm tor}$ (n)  & 0.30 (Y) & 0.33 (Y)\\ 
& $\phi_{\rm tor}$ (n+1) & 0.96 (Y) & 0.96 (Y)\\
    & $\phi_{\rm tor}$ (n+2)& 0.36 (Y) & 0.42 (Y)\\
    & $\phi_{\rm tor}$  (n+3)& 0.31 (Y) & 0.35 (Y)\\
\cline{1-4}
4.0 &$\phi_{\rm tor}$ (n) & 0.03 (N) & $-$0.08 (N)\\
& $\phi_{\rm tor}$ (n+1) & 0.92 (Y) & 0.91 (Y)\\
    & $\phi_{\rm tor}$ (n+2) & 0.14 (N) & 0.07 (N)\\
    & $\phi_{\rm tor}$ (n+3) & 0.12 (N) & 0.18 (N)\\
\cline{1-4} 
\end{tabular}
\label{table3}
\end{table}
Let us now discuss the physics behind these correlations. Our discussion 
will be largely based on \citet{JCC07, YNM08, CB11}. 
In  \bl\ models, the toroidal field 
gives rise to the poloidal field through $\alpha$ effect 
(not the mean-field $\alpha$ but a simplified parameterization of \bl\ process). 
In this process, there is nonlinearity and 
some randomness. The poloidal field generated in the 
low latitudes is first transported to high latitudes and then to the deep CZ
through meridional circulation and diffusion. 
Finally, this poloidal field gives rise to the toroidal field for the 
following cycle through differential rotation. 
Thus the dynamo chain can be written as follows\\
$\phi_{\rm tor} (n) \xrightarrow[\text{+ Nonlinearity}]{\text{Randomness}} \phi_{\rm r} (n)  \xrightarrow[\text{Deterministic}]{(\Omega~{\rm effect})} \phi_{\rm tor} (n+1)...
$
\\
Clearly, the correlation between the $\phi_{\rm r} (n)$  and $\phi_{\rm tor} (n)$
is affected by randomness and nonlinearity. 
When the model  operates near the critical $\alpha_0$, the nonlinearity in the $\alpha$
is weak and  
thus, the correlation 
between $\phi_{\rm r} (n)$ and $\phi_{\rm tor} (n)$ is determined only by the randomness included 
in the \bl\ $\alpha$. The randomness that
we have included in our model tries to reduce this correlation and 
thus the correlation between $\phi_{\rm r} (n)$ and $\phi_{\rm tor} (n)$ is not perfect 
even at the smallest $\alpha_0$ ($\hat{\alpha}_0 = 2.0$; \Tab{table1}). 
We emphasize that as long as $\alpha_0$ is not completely random, there is  
a significant correlation between $\phi_{\rm r} (n)$ and $\phi_{\rm tor} (n)$. 
This is what is seen in \Tab{table1} for $\hat{\alpha}_0 = 2$ in Models I and II.

As the dynamo becomes supercritical, the nonlinearity increases. 
This nonlinearity along with the randomness spoils the linear relation in the chain: 
$\phi_{\rm tor} (n) \rightarrow \phi_{\rm r} (n)$. Hence, 
the correlation between $\phi_{\rm r} (n)$ and $\phi_{\rm tor} (n)$ reduces with the increase of $\alpha_0$. 
As seen in \Tab{table1}, simulations at $\hat{\alpha}_0 > 3$ do not show this correlation strongly. 

Next, the strong correlation between $\phi_{\rm r} (n)$ and $\phi_{\rm tor} (n+1)$ 
in all the runs are easy to understand.
As $\phi_{\rm r} (n)$ $\rightarrow$  $\phi_{\rm tor} (n+1)$ involves 
only $\Omega$ effect which is deterministic 
in our model,
the correlation between $\phi_{\rm r} (n)$ and $\phi_{\rm tor} (n+1)$ must be strong 
in all runs. We note that this correlation holds in any model, 
as long as the poloidal field feeds back 
to the toroidal one through diffusion and/or advection \citep{CB11}. 
This is what is seen in \Tab{table1}.
There is a slight decrease of this $n$--$(n+1)$ correlation with the increase of
$\alpha_0$. This is possibly due to the systematic decrease of mean cycle period 
with the increase of $\alpha_0$; see \Tab{table2}. When the cycle period
decreases, the transport time of the poloidal field effectively decreases.
Thus, the less efficient transport of poloidal field causes to decrease
the correlation between the polar flux and the next cycle toroidal flux.

Finally, the $n$--$(n+2)$ and $n$--$(n+3)$ correlations are linked to the part: 
$\phi_{\rm tor} \rightarrow  \phi_{\rm r}$. If the nonlinearity is 
weak, then the correlations of $\phi_{\rm r} (n)$ with $\phi_{\rm tor} (n+2)$ 
and $\phi_{\rm tor} (n+3)$ will be determined by the randomness
in $\alpha$ and the efficiency of the magnetic field transport 
and hence there will be some correlation. 
As the dynamo is marginally 
supercritical at $\hat{\alpha}_0=2$, multiple cycle correlations are seen in top four panels of \Fig{fig:dicorr}.
On the other hand, in the supercritical regime, these multiple cycles
correlations will disappear because the chain $\phi_{\rm tor} \rightarrow \phi_{\rm r}$ is spoiled by
the nonlinearity and the randomness. Thus all $n$--$(n+2)$ and $n$--$(n+3)$ correlations are negligible 
at large $\hat{\alpha}_0$; see \Tab{table1} and the bottom four panels of \Fig{fig:dicorr}.

In \Fig{fig:ModelI} we observe that when the dynamo is near the critical, it produces extended episodes of weaker activity, resembling the solar grand minima, and as the
supercriticality increases, the frequency of grand minima decreases. In highly supercritical regime, the model does not produce any grand minima even when the fluctuation level is increased (because the dynamo growth is large and the magnetic field quickly grows from the weaker value). During these grand minimum phases, the model becomes very linear and causes multi-cycle correlations in the polar field. In the supercritical regime, as the model does not enter into any extended grand minimum phase, the model remains nonlinear all the time and thus it cannot produce multi-cycle correlation.
In fact, we have seen that if we exclude these grand minimum phases from the data of the weakly-supercritical regime (red colored zones in \Fig{fig:ModelI}), then we observe that the multi-cycle correlation is reduced to one cycle.
This supports our conclusion that when the dynamo is marginally supercritical (weakly nonlinear) and when the $\alpha_0$ is not completely random, 
there will be a good correlation between $\phi_{\rm tor} (n)$ and $\phi_{\rm r} (n)$.
When this $n$--$n$ correlation holds, the dynamo chain will allow 
$\phi_{\rm r} (n)$--$\phi_{\rm tor} (n+2)$ and $\phi_{\rm r} (n)$--$\phi_{\rm tor} (n+3)$
correlations to persist.


We recall that both the simulations labeled as the diffusion-dominated and advection-dominated regimes of \citet{YNM08}
were performed at $\alpha_0 =30$~\mps (their  (Run 1 and 2). They claimed that the one-cycle correlation found in diffusion-dominated simulation
is due to the dominant contribution of the diffusion
over the advection. We do not agree with this conclusion 
because if this is the case, then at smaller $\alpha_0$ also the same correlations
would have been observed,
which is not the case; see \Tab{table1} and \Fig{fig:dicorr}.
In reality,
the super-criticality of their two models are different.
As we have shown (\Fig{fig:dynamo}) that the critical $\alpha_0$ in their two models are 6.2 and 10~\mps. 
Hence, $\alpha_0 =30$~\mps\ correspond to $\hat{\alpha}_0=4.8387$ in Model~I (their diffusion-dominated regime)
and $\hat{\alpha}_0=3$ in Model~II (their advection-dominated regime). Thus, what \citet{YNM08} call diffusion-dominated is actually highly supercritical 
than their advection-dominated model. 
And we have seen in our study that with the increase of super-criticality, 
the model becomes more nonlinear (grand minima becomes very rare; see \Sec{sec:appendix} and \Fig{fig:suppli}) 
and  the multiple cycles correlations are reduced
to one cycle.

Incidentally, \citet{Hazra20} did not find multiple cycle correlations even in their advection-dominated model
(see insignificant correlation values for $n$--$n$, $n$--$(n+2)$ and $n$--$(n+3)$ as given in their Tables 1 and 2, last columns)---a clear contradiction to \citet{YNM08}.
The reason for not finding multiple-cycle correlations in \citet{Hazra20} is that they included a mean-field $\alpha$, 
in addition to the \bl\ source for the poloidal field and this additional $\alpha$ possibly made the model considerably supercritical.
And, as we have shown that in the supercritical regime, even if the dynamo is advection-dominated, the polar field memory 
is limited to only one cycle.
 
\citet{KN12} showed that when a downward turbulent pumping in the poloidal field is included in the advection-dominated
dynamo, the multiple-cycle correlation is reduced to one cycle. Actually, when pumping is included in the model,
the dynamo becomes strong (because of the suppression of magnetic flux through the surface; see \citet{Ca12,KC16,KM18}).
And the dynamo transition happens at smaller $\alpha_0$. Thus, the same model which produced multiple cycle correlations
because of weakly super-criticality becomes heavily supercritical with the inclusion of turbulent pumping. 
This is the hidden reason for displaying one cycle correlation in the advection-dominated model with turbulent pumping.

We do not mean that the turbulent transports, namely, the turbulent diffusivity, pumping and meridional flow 
do not play any role in determining these correlations.
The transport is necessary to connect the spatially segregated source regions of the poloidal
and the toroidal fields. Through various transport processes, the polar field of a cycle ($\phi_{\rm r} (n)$)
is strongly correlated with the amplitude of next cycle toroidal field ($\phi_{\rm tor} (n+1)$).
The generation of poloidal field and thus the connection between $\phi_{\rm tor} (n)$ and $\phi_{\rm r} (n)$
is also possible through turbulent transport. However, as we do not change any parameter of diffusion and meridional flow in these models, our correlations are not affected by turbulent transport processes.

\begin{figure*}
\centering
\includegraphics[scale=0.4]{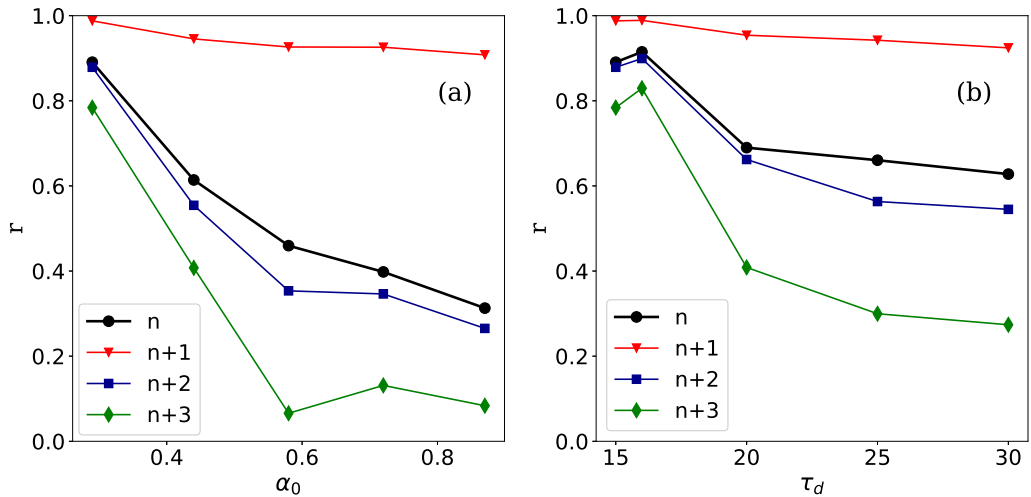}
\caption{
Results from the time delay dynamo model: 
Variations of the correlation coefficients of the peak poloidal fields of cycle $n$ with the peak toroidal fields of cycle 
$n$ (black line), $n+1$ (red), $n+2$ (blue), and $n+3$ (green)  with 30$\%$ fluctuations.
In the left panel, $\tau_d$ is fixed at $15$ and $\alpha_0$ is increased, 
while in right panel, $\alpha_0$ is fixed at $0.29$ and $\tau_d$ is increased.
}
\label{fig:delay}
\end{figure*}

\begin{figure}
\centering
\includegraphics[scale=0.4]{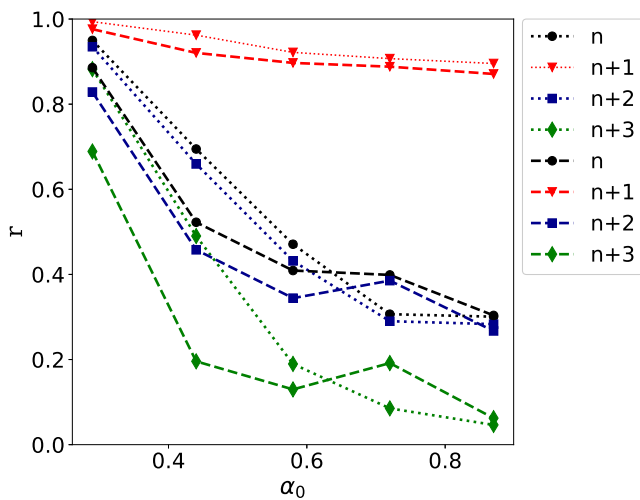}
\caption{Same as \Fig{fig:delay}(a) but at higher $\alpha$ fluctuations; dotted and dashed for 45$\%$ and 100$\%$ fluctuations, respectively.}
\label{fig:delay1}
\end{figure}

%

Now we examine the results from Models~III and IV (\Tab{table1}).
These models produce cycles even at $\hat{\alpha}_0 = 1$. This is possibly due to
much smaller diffusivity and different  $\alpha$ prescription. 
We observe that the $n$--$n+1$ cycle correlation is very strong in all the runs.
For Models I and II, this was not the case, because the diffusivity of the 
poloidal field in the CZ was about 20 times higher than that in Models III and IV.
This higher diffusion in Models I and II allowed some of the fluctuations in the $\alpha$ to 
propagate to the same cycle toroidal flux. 
Despite these differences, we find similar behavior with the increase of $\hat{\alpha}_0$.
That is, when the dynamo is only
marginally supercritical ($\hat{\alpha}_0$ near 1), multiple cycle correlations exist. On the other hand,
when the dynamo is supercritical, only one cycle correlation holds. 
This conclusion remains unaffected even when we increase the level of fluctuations. \Tab{table3} shows the results at 150$\%$ and 200$\%$
fluctuation levels. If we increase the fluctuation levels too much, then the model tends to decay, particularly Models I and II in the weakly supercritical regime.

%
%
We note that even the model without meridional flow: Model~V also shows the same behaviour of the polar field with the increase of the supercriticality; see \Tab{table1}. This confirms that the advection plays no role in setting the memory of polar field beyond one cycle which is in contradiction to \citet{YNM08}.

Finally, we observe the results from Model~VI. As noted earlier, the set of runs in Model~VI, is the same as that in Model~III, except $\alpha_0$ 
is fixed at 0.4~\mps\ and $\eta_0$ is decreased (or $\hat{\eta}_0 = \eta_0^{\rm crit}/\eta_0$ is increased) 
in each run.
When $\eta_0$ is decreased, the model becomes less diffusion dominated. Interestingly, even when the model
becomes less diffusive, we observe the shorten of memory of the polar flux.
In \Tab{table1}, we observe that at smaller $\hat{\eta}_0$ (1 and 1.5) multiple cycle correlations hold, while at 
larger $\hat{\eta}_0$ only $n$--$n+1$ correlation holds. The correlation plots for marginally supercritical ($\hat{\eta}_0 = 1$)
and highly supercritical ($\hat{\eta}_0 =4$) are shown in \Fig{fig:etavaried}.

\section{Results from Time Delay Dynamo Model}
\label{sec:res2}
To demonstrate that the above conclusion is not limited to only the flux transport dynamo models, but also generic in a simplified
\bl\ type dynamo models, we present the results from a truncated low-order time delay model. 
As discussed in \Sec{sec:moddelay},
the (physically-motivated) time delay between poloidal and toroidal fields are captured by the parameters $T_0$ and $T_1$.
In the same manner, as we have done for the flux transport dynamo models, we include stochastic fluctuations in the poloidal source; however, in this delay model, we first include $30\%$ fluctuations in $\alpha_0$, appearing in \Eq{eq:delay2}. 
We note that previous study based on the same model \citep{Ha14} also includes this amount of fluctuations in $\alpha_0$.
The results for $45\%$ and $100\%$ fluctuations are shown in \Fig{fig:delay1}.
Again in this model, we find that, as we make the dynamo more and more super-critical by increasing $\alpha_0$ (\Fig{fig:delay}a and \Fig{fig:delay1})
or diffusion time $\tau_d$ (\Fig{fig:delay}b), correlations: $n$--$n$, $n$--$n+2$ and $n$--$n+3$ all are diminished and only the $n$--$n+1$ correlation
is retained.

\section{Conclusions and Discussions}
\label{sec:conc}

By performing stochastically forced dynamo simulations at different parameter regimes and with
different models, we identify the memory of the polar flux, explicitly, 
how the peak polar flux of cycle $n$, $\phi_{\rm r} (n)$ is correlated with the toroidal flux of the same cycle ($\phi_{\rm r} (n)$)
and the subsequent cycles. We show that when the dynamo is barely above the transition 
or when the dynamo is not heavily supercritical, the nonlinearity in the \bl\ $\alpha$ is weak. 
In this case, the
$\phi_{\rm tor} (n) \rightarrow \phi_{\rm r} (n)$ chain is only affected by the fluctuations in $\alpha$ and 
as long as $\alpha$ is not completely stochastic,
the polar flux is linearly correlated with the toroidal flux of the same cycle
and many subsequent cycles. 
The dynamo in this weakly supercritical regime
also produces occational grand minima.
On the other hand, when the dynamo is sufficiently above the critical dynamo transition,
the nonlinearity and stochasticity in $\alpha$ spoil the linear correlation between 
$\phi_{\rm tor} (n)$ and $\phi_{\rm r} (n)$. 
This subsequently breaks all multiple-cycle correlations: $\phi_{\rm r} (n)$ vs $\phi_{\rm tor} (n)$, $\phi_{\rm r} (n)$ vs $\phi_{\rm tor} (n+2)$ and $\phi_{\rm r} (n)$ vs $\phi_{\rm tor} (n+3)$, and only the correlation $\phi_{\rm r} (n)$ vs $\phi_{\rm tor} (n+1)$ survives.


In the flux transport dynamo models, the polar flux is coupled to the toroidal flux of the following cycle through the meridional circulation and turbulent diffusivity. As long as this coupling is there, the polar flux is highly correlated
with the following cycle toroidal flux; see \citet{CB11} for detailed study. 
This coupling is indeed observed in terms of the correlation between the surface polar flux and following cycle amplitude \citep{CCJ07, WS09, Kit18, KMB18, Pawan21}. 
In our study, we have shown that when the dynamo is highly supercritical,
the memory of the polar flux cannot be propagated to more than one cycle. This one-cycle memory is independent of the relative importance of the diffusion in the model. 
We do not agree with the explanation given in previous studies \citep{YNM08,KN12,Hazra20} 
which say that the multi-cycle memory of polar field is determined by the relative importance of advection vs diffusion or turbulent pumping; see \Secs{sec:rescorr}{sec:appendix} for details.

Analyses of the polar faculae counts \citep{Muno13}
show that
the polar flux at the cycle minimum has no statistically significant correlation beyond one cycle. This however does not reflect the true feature of the solar cycle because of two reasons.
First, the polar faculae data is too noisy and poorly binned. Hence, the weak multi-cycle correlation which might be present in real sun is not detected in this poor quality data. 
Second, the observed polar faculae data is available only for ten cycles (1907--2011) when the solar activity was relatively strong and it does not cover any extended weak phase like the Maunder or Dalton minima. Basically, what we want to point out is that the observed data only represent a small subset of the whole solar cycle pattern. For example, if we take a small subset of the data presented in \Fig{fig:ModelI}, then we get a different result than what we get from the full data set. If we exclude the extended weaker field episodes like grand minima (marked by red color), then the multi-cycle correlation is reduced to one cycle. When the model produces extended weak cycles, the nonlinearity becomes weak and the multi-cycle correlation of the polar field is unavoidable. 
As the sun produces frequent grand minima and 
these are produced only when the dynamo is weakly supercritical, we can indirectly say that the solar dynamo is weakly supercritical. We need better and longer data of polar field to confirm the multi-cycle correlation that is the outcome of weakly supercritical dynamo.
Due to the simplicity (e.g., ignoring the back reaction of the magnetic field on the flow and turbulent transport), 
and uncertainties of some parameters
in our models, we, however, cannot provide the exact value of super-criticality of the solar dynamo from our simplified simulations. 
Thus our prediction is in qualilative agreement with the previous studies \citep{KN17} which suggested that solar dynamo is weakly supercritical. 


\acknowledgements
Authors sincerely thank Dibyendu Nandi, Kristof Petrovay, Arnab Rai Choudhuri, and Leonid Kitchatinov for reading this manuscript and providing valuable comments and finding a few errors;  all these helped to improve the presentation of results. 
Authors also gratefully acknowledge the constructive comments and suggestions from an anonymous referee.
Financial supports from Department of Science and Technology (SERB/DST), India 
through the Ramanujan Fellowship (project no SB/S2/RJN-017/2018) and 
ISRO/RESPOND (project no SRO/RES/2/430/19-20) are acknowledged.
VV acknowledges financial support from DST through INSPIRE Fellowship.
BBK acknowledges the funding provided by the Alexander von Humboldt Foundation.

\appendix
\label{sec:appendix}

\section{Additional material to demonstrate the problem in the conclusion of \citet{YNM08}}

\begin{figure*}
\centering
\includegraphics[scale=0.3]{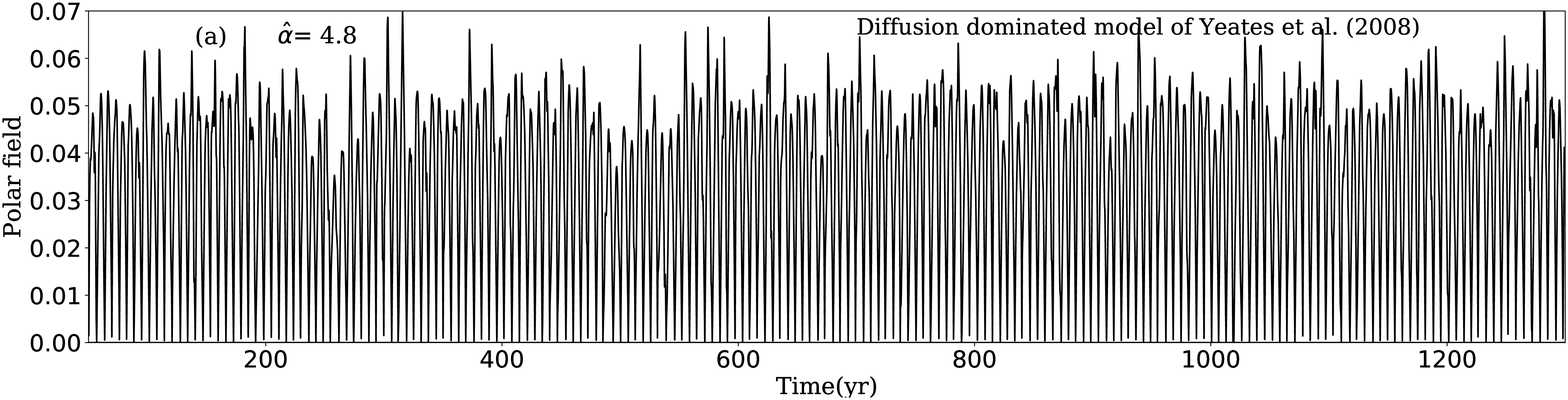}
\includegraphics[scale=0.3]{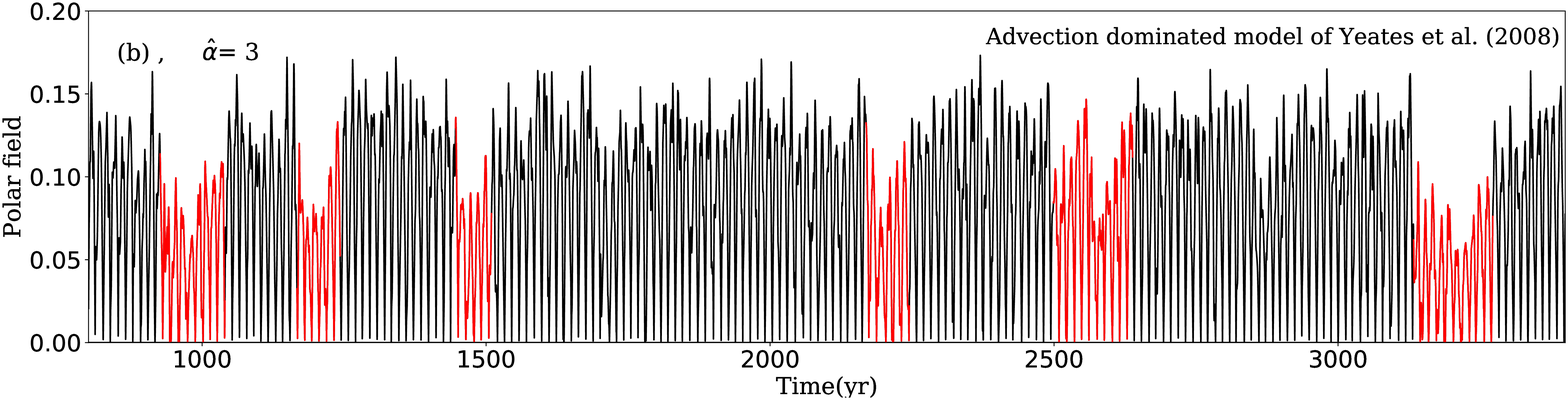}
\caption{
Top and bottom: Time series of surface polar flux from Models I and II at $\alpha_0 = 30$ \mps\ (or $\hat{\alpha}=4.8$ and 3) which respectively correspond to the diffusion- and advection-dominated models of \citet{YNM08}.
}
\label{fig:suppli}
\end{figure*}

\strut
\Fig{fig:suppli} top and bottom panels demonstrate the time series of polar flux of runs at $\alpha_0 = 30$ \mps\ in Models I and II, respectively. We note that $\alpha_0 = 30$ \mps\ corresponds to $\hat{\alpha} = 4.8$
and 3 in these two runs. \citet{YNM08} called these two runs as diffusion- and advection-dominated. Their diffusion-dominated model (shown in the top panel of \Fig{fig:suppli}) indeed has a larger supercriticality, while the advection-dominated model (lower panel) has a lower supercriticality, and thus it produces several grand minimum-like weaker activity episodes. 
The model 
is close to a linear one when it is in these grand minimum phases and thus it produces multi-cycle correlations. 
These multi-cycle correlations are not due to the dominance contribution of the advection over the diffusion in the model as suggested by \citet{YNM08}.

\clearpage

\bibliography{paper}
\bibliographystyle{aasjournal}



\end{document}